\documentstyle[aps]{revtex}

\begin{document}
\draft
\title{Fermi Liquid Damping and NMR Relaxation in Superconductors}
\author{S. Tewari and J. Ruvalds}
\address{Physics Department\\
University of Virginia\\
Charlottesville, VA 22903}
\date{\today }
\maketitle

\begin{abstract}
Electron collisions for a two dimensional Fermi liquid (FL) are shown to
give a quasiparticle damping with interesting frequency and temperature
variations in the BCS superconducting state. The spin susceptibility which
determines the structure of the damping is analyzed in the normal state for
a Hubbard model with a constant on--site Coulomb repulsion. This is then
generalized to the superconducting state by including coherence factors and
self energy and vertex corrections. Calculations of the NMR relaxation rate
reveal that the FL damping structure can reduce the Hebel--Slichter peak, in
agreement with data on the organic superconductor (MDT-TTF)$_2$AuI$_2$.
However, the strongly suppressed FL damping in the superconducting state
does not eliminate the Hebel-Slichter peak, and thus suggests that other
mechanisms are needed to explain the NMR data on (TMTSF)$_2$ClO$_4$, the
BEDT organic compounds, and cuprate superconductors. Predictions of the
temperature variation of the damping and the spin response are given over a
wide frequency range as a guide to experimental probes of the symmetry of
the superconducting pairs.
\end{abstract}


\section{ Introduction}

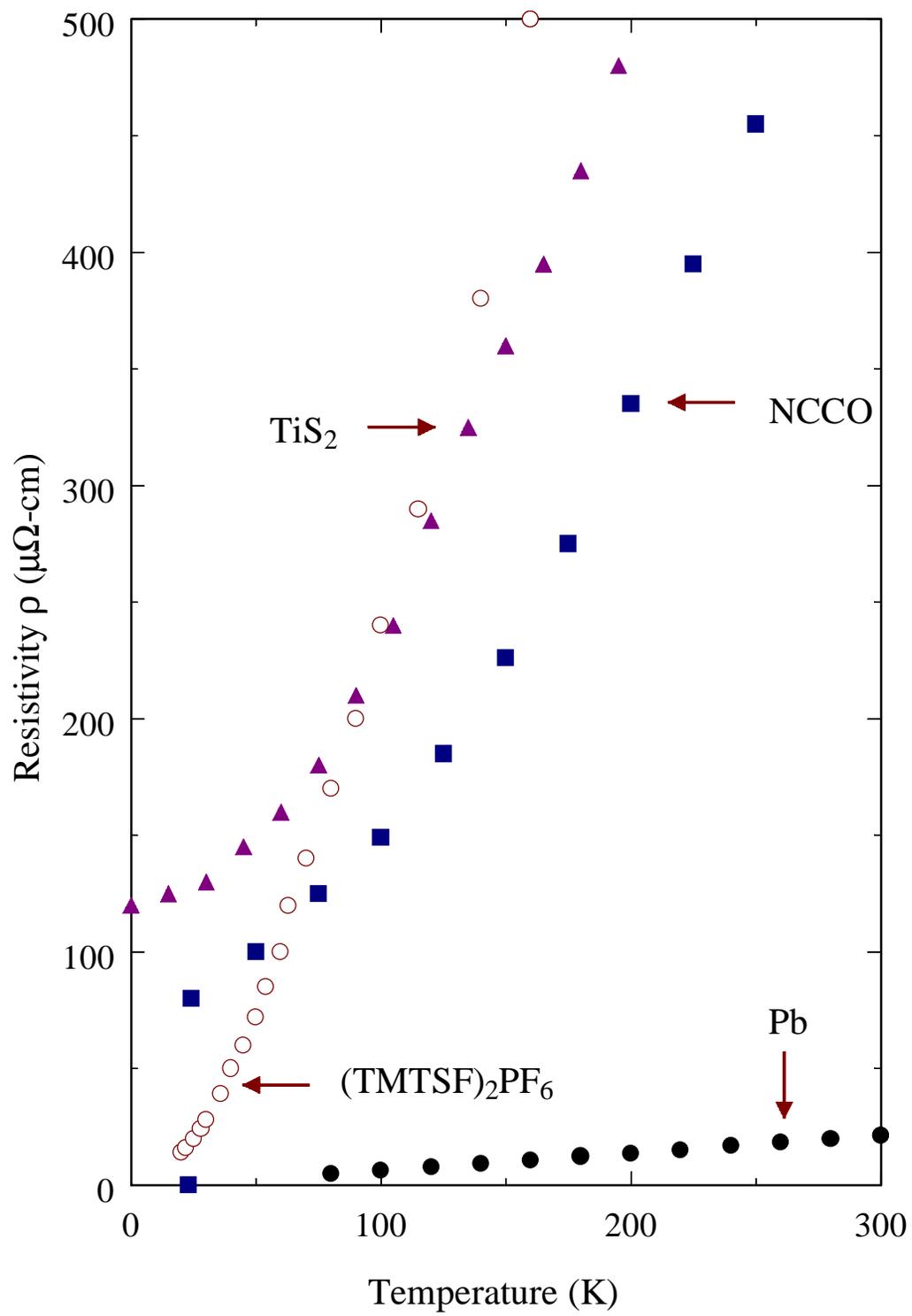
\begin{figure}
\caption{The resistivity as a function of temperature is shown for various
materials at temperatures above the superconducting transition. Note that
the resistivity of Pb at these temperatures is linear in $T$ (above the
Debye temperature) and is much lower than that of the organic metal (TMTSF)$%
_2$PF$_6$ and the layered compounds TiS$_2$ and Nd$_{1.84}$Ce$_{0.16}$CuO$%
_{4-y}$, whose resistivities show a quadratic dependence on temperature.
This $T^2$ variation resembles Fermi Liquid damping with an anomalous
enhancement of the electron-electron scattering.}
\label{rho}
\end{figure}

The damping of quasiparticles via electron collisions is expected to produce
a $T^2$ variation of the resistivity in a standard Fermi liquid described by
a spherical Fermi surface. In ordinary metals this contribution is so weak
that it is hardly detectable\cite{ashcroft}. However, recent discoveries of
a dominant $T^2$ resistivity contribution in several anisotropic metals such
as organic superconductors and layered cuprate and sulfide alloys provide an
interesting challenge from the theoretical point of view since the observed
magnitudes of such resistivities are very large. Examples\cite{data,julien}
of the unconventional resistivities are shown in Figure \ref{rho}.

In view of the conventional wisdom regarding the expected weakness of the
electron-electron scattering contribution to the resistivity, alternate
interpretations of the TiS$_2$ data were proposed on the basis of unusual
phonon scattering. However, decisive evidence for the Fermi liquid origin of
the scattering was discovered by Julien\cite{julien} in the infrared spectra
of TiS$_2$. These spectra show a remarkable frequency and temperature
variation in accordance with the Luttinger damping for a Fermi sphere that
was derived to all orders of perturbation theory for the Coulomb interaction%
\cite{luttinger}. The huge value of the resistivity of TiS$_{2\text{ }}$is
on par with the data for the Nd$_{2-\text{x}}$Ce$_{\text{x}}$CuO$_4$ cuprate
which also has a layered structure. It is also comparable to the resistivity
of many organic metals such as that of (TMTSF)$_2$PF$_6$ shown in Figure \ref
{rho}. The organics have quasi-two dimensional Fermi surfaces with a high
degree of anisotropy. By contrast, the much smaller resistivity of lead
shown in Figure \ref{rho} provides an example of the scale set by strong
electron-phonon contributions.

The purpose of the present work is to calculate the quasiparticle damping
for a two dimensional electron system due to spin fluctuation scattering
arising from the constant on-site Coulomb interaction $U$ in a Hubbard
model. We first compute the standard normal state damping and then focus on
the influence of an isotropic superconducting energy gap on the frequency
and temperature variation of the self energy. The superconducting energy gap
is expected to reduce the damping in a characteristic way that depends on
the symmetry of the order parameter as well as the source of the damping.
The spin susceptibility, which enters in the calculation of the damping, is
first analyzed for the normal electrons and then generalized to include the
BCS coherence factors in the superconducting state. Our goal is to use the
computed damping in a calculation of the NMR relaxation rate, with
particular interest in the superconducting state response. The relaxation
rate is obtained from a momentum average of the spin susceptibility that
includes the calculated FL self-energy corrections. Vertex corrections to
the susceptibility are also calculated and shown to be small in the regime
considered in the NMR studies.

We apply our results to experimental NMR measurements on the organic
superconductors. These provide insight into the structure caused by the
damping as well as the symmetry of the superconducting state. The
modification of the Hebel--Slichter (HS) peak is found to be particularly
instructive in the case of the anomalies seen in the organics, although the
general phenomenon is relevant to the high temperature cuprate
superconductors as well.

Luttinger\cite{luttinger} originally derived the classic three dimensional
Fermi liquid property of a quasiparticle damping arising from electron
collisions that vanishes at the Fermi energy at zero temperature and then
follows a quadratic variation in temperature $T$ and frequency $\omega $
(measured from the Fermi energy). Hodges {\it et al.}\cite{wilkins} and
others\cite{quinn} calculated the quasiparticle damping and resistivity for
a cylindrical Fermi surface, which introduces logarithmic corrections to the
damping at low $T$ and $\omega $.

We first calculate the two dimensional Fermi liquid damping in the normal
state by performing a momentum average over the exact spin susceptibility
for a non-interacting electron gas. We then proceed to compute the
susceptibility as well as the self energy in the superconducting state, and
probe the influence of the calculated damping on the NMR response which
involves a momentum--averaged spin susceptibility.

The original motivation for our present work was the mysterious absence of a
Hebel--Slichter peak in the NMR response of high temperature cuprate
superconductors\cite{cuprates}. The resistivity of the optimally doped
cuprates above the superconducting transition is typically linear in
temperature in contrast to the Fermi liquid behavior considered here. A
similar physical origin of electron-electron scattering in both cases may be
plausible, because alloying changes the linear $T$ to a $T^2$ variation in
many of the cuprates\cite{takagi}. Historically the HS peak in the NMR
relaxation is considered to be one of the key successes of the BCS theory
since it occurs in various ordinary metals\cite{hebel}. The organic
superconductors yield cases where the HS peak appears, (MDT-TTF)$_2$AuI$_2$
\cite{takahashi}, and also examples like (TMTSF)$_2$ClO$_4$\cite{takigawa}
where the peak is suppressed. The NMR relaxation in the BEDT\cite{kanoda}
compounds resembles the cuprate anomalies, such as a $T^3$ behavior at low $%
T $ in the superconducting state. Hence the present analysis is directed at
the organics which often have the quadratic $T$ variation of the resistivity
that is compatible with our Fermi liquid analysis.

The absence of the coherence peak has attracted considerable previous
theoretical interest. Hasegawa and Fukuyama\cite{hasegawa} derived the NMR
response for s--wave and other symmetries of the energy gap appropriate to
organic superconductors. They obtained weaker, but nonetheless finite, HS
peak structure for both singlet and triplet pairing states in which the
order parameter exhibits line nodes on the Fermi surface. Their calculation
did not include the quasiparticle damping.

Other groups have examined the effects of a strong enhancement of the
quasiparticle damping near $T_c$, which broadens the BCS singularity in the
superconducting density of states. Some theory groups have invoked a variety
of phenomenological models for the damping as a function of temperature\cite
{damptheory}. Typically, these authors use a power law variation for the $T$
dependence of the damping (with a $T^3$ behavior as one example) and neglect
the frequency variation. Phenomenological models have been proposed to fit
the observed spin dynamics in the cuprates\cite{pines}. The ``marginal''
Fermi liquid response hypothesis has been used to generate a damping model
which has also been applied to the NMR measurements\cite{varma}.

Tight binding energy band models have also been investigated in the context
of spin dynamics of cuprates by Bulut and Scalapino\cite{bulut} and by
Levin's group\cite{levin}. The NMR relaxation for s and d--wave energy gaps
has been calculated\cite{damptheory,bulut} using phenomenological models for
the damping. The quasiparticle damping from spin fluctuations in the
superconducting state has also been independently computed within a tight
binding model\cite{quinlan}.

Our analysis of the influence of the damping on the NMR spectra in the
superconducting state is based on a standard approximation for the
momentum--averaged spin susceptibility. This allows analytic expressions to
be derived and used in the calculation of the damping, thus reducing the
number of numerical integrations. However, this limits the present approach
to reasonably isotropic Fermi surfaces, and cannot account for features such
as nesting.

Phonon damping has furnished yet another more traditional explanation for
the suppressed HS peak, provided that a large electron-phonon coupling
combined with a small Coulomb pseudopotential are assumed\cite{allen}.
Phonon contributions to the resistivity may be one distinguishing feature of
the mechanism responsible for the NMR anomalies, and the organic resistivity
data in Figure \ref{rho} provide a challenge in this regard since the $T^2$
variation is difficult to reconcile with phonon scattering.

Considering the wide array of theoretical proposals for the NMR anomalies,
it is important to make distinctions based on specific features such as the
response over a wide range of temperature, and consistency of parameters
with other experimental clues. The $T^2$ damping which characterizes the
organic superconductors considered here is one example. The temperature
dependence of the NMR relaxation rate at low $T$ is also a primary
constraint on theory since it is particularly sensitive to the symmetry of
the order parameter, while the HS peak structure is sensitive to the damping
as well as the gap symmetry. Thus our calculations provide insight into the
role of damping in the organic superconductors and the NCCO cuprate which
also exhibits a $T^2$ resistivity above the superconducting transition.

A theoretical basis for the anomalous linear T variation of the resistivity
that is an ubiquitous feature in all of the optimally doped high temperature
superconductors discovered so far is the nested Fermi Liquid theory (NFL)%
\cite{nfl}. The corresponding calculations for the NFL damping in the
presence of an isotropic energy gap by Rieck {\it et al.}\cite{rieck} reveal
a very dramatic suppression of the nested spin susceptibility in the
superconducting state which strongly reduces available scattering states for
electron collisions. Hence the NFL damping is greatly reduced at frequencies
lower than thrice the energy gap $\Delta $. The latter damping structure is
compatible with microwave surface resistance data on the YBCO superconductor
in the vicinity of $T_c$ where the damping drops by four orders of magnitude
within a few degrees of $T_c$. There are also numerous earlier calculations
that found a sharp suppression of the damping in the case of
phenomenological models that are extensions of the ``marginal'' Fermi liquid
hypothesis\cite{varma} for the susceptibility. These are discussed in
reference\cite{rieck}.

Recently Won and Maki\cite{maki} have shown that a nesting model yields NMR
relaxation without a HS peak if indeed the scattering processes near the
nesting vector form the dominant contribution to the susceptibility.

We develop the formalism for the susceptibility and self energy in section
II, and present the calculated Fermi liquid damping results. The NMR
relaxation is discussed in section III, and the conclusions of our study are
in section IV.

\section{Formalism}

We consider the Hubbard Hamiltonian
\begin{equation}
H=\sum_{{\bf k},\sigma }\epsilon _{{\bf k}}c_{{\bf k},\sigma }^{\dagger }c_{%
{\bf k},\sigma }^{}+U\sum_{{\bf p,q,k}}c_{{\bf p+q},\uparrow }^{\dagger }c_{%
{\bf p},\uparrow }^{}c_{{\bf k-q},\downarrow }^{\dagger }c_{{\bf k}%
,\downarrow }^{},  \label{hamil}
\end{equation}
where $\epsilon _{{\bf k}}$ is the energy of an electron in two dimensions,
and $c_{{\bf k},\sigma }^{\dagger }$ and $c_{{\bf k},\sigma }^{}$ are the
electron creation and destruction operators. The constant on-site Coulomb
repulsion $U$ restricts the scattering to spin fluctuations. Within the Born
approximation considered here, the dominant effect is from the
non-interacting spin susceptibility and the cross section is proportional to
$U^2$. In other words the self energy requires a momentum sum over $U^2\chi(%
{\bf q},\omega)$. Higher order RPA corrections will naturally enhance the
cross section, but are not included here.

\subsection{Normal State}

The spin susceptibility in the normal state for non-interacting electrons is
\begin{equation}
\chi \left( {\bf q},\omega \right) =\sum_{{\bf k}}\frac{f\left( \epsilon _{%
{\bf k+q}}\right) -f\left( \epsilon _{{\bf k}}\right) }{\omega -\epsilon _{%
{\bf k+q}}+\epsilon _{{\bf k}}+i\Gamma }  \label{susc}
\end{equation}
where $f(\epsilon )$ is the Fermi function. Within the leading order Born
approximation, the imaginary part of the self energy arising from
electron-electron scattering is\cite{agd}
\begin{equation}
\Gamma \left( {\bf k},\omega \right) =\frac{U^2}2\int d\omega ^{\prime
}\left[ \coth \left( \frac{\omega ^{\prime }}{2T}\right) -\tanh \left( \frac{%
\omega ^{\prime }-\omega }{2T}\right) \right] \int \frac{d{\bf q}}{\left(
2\pi \right) ^3}\chi ^{\prime \prime }\left( {\bf q},\omega \right) \delta
\left( \omega -\omega ^{\prime }-\epsilon _{{\bf k+q}}\right) ,
\label{gamma}
\end{equation}
where $\chi ^{\prime \prime }$ is the imaginary part of $\chi ({\bf q}%
,\omega )$.

We first verified numerically that the imaginary part of the susceptibility
in Eq. (\ref{susc}) does not vary significantly with temperature for a
normal two-dimensional electron gas, and hence one may use the zero
temperature analytic result in calculating $\Gamma $. Defining the frequency
$\omega $ in units of $\hbar k_F^2/m$ and the wavevector $q$ in units of the
Fermi wavevector $k_F$, the $T=0$ analytic form for $\chi ^{\prime \prime }$
is
\begin{equation}
\chi ^{\prime \prime }\left( {\bf q},\omega \right) =\frac{N(0)}q\left\{
\theta \left[ 1-\left( \frac \omega q-\frac q2\right) ^2\right] \left[
1-\left( \frac \omega q-\frac q2\right) ^2\right] ^{\frac 12}-\theta \left[
1-\left( \frac \omega q+\frac q2\right) ^2\right] \left[ 1-\left( \frac
\omega q+\frac q2\right) ^2\right] ^{\frac 12}\right\}  \label{imchi}
\end{equation}
where $N(0)$ is the density of states at the Fermi level, and $\theta (x)$
is the Heaviside function that is unity for $x>0$ and zero otherwise. Hodges%
\cite{wilkins} and others\cite{quinn} have used the above analytic form to
calculate the damping in the asymptotic limits of small $T$ and $\omega $,
by performing the momentum integrations (Eq. \ref{gamma}) to obtain the
transport lifetime. Note that the imaginary part of the self energy is $%
\Gamma =[1-f(\omega )]^{-1}\hbar /\tau $ where $\tau $ is the quasiparticle
transport lifetime\cite{mahan}.

To obtain a damping over the entire range of frequency and temperature, we
employ an alternate approximation: the susceptibility in Eq. (\ref{gamma})
is first averaged over momentum and gives at low frequencies ($\omega
<\epsilon _F$),
\begin{equation}
\left\langle \chi ^{\prime \prime }\left( {\bf q},\omega \right)
\right\rangle _q=\frac{\pi N(0)}W\omega  \label{chifl}
\end{equation}
where $W$ is the energy bandwidth, and $N(0)$ is the density of states at
the Fermi energy. We note that the momentum-averaged susceptibility is
independent of temperature for the Fermi liquid. Replacing this average in
Eq. (\ref{gamma}), we evaluate $\Gamma $ numerically and find a
quasiparticle damping that is quadratic in frequency and temperature. The
limiting cases are
\begin{eqnarray}
\Gamma (\omega &=&0)=BT^2 \\
\Gamma (T &=&0)=C\omega ^2
\end{eqnarray}
where $B=$ $U^2N^2(0)\pi ^3/2W$ and $C=U^2N^2(0)\pi /2W$. The present method
does not yield the $T^2\ln T$ dependence found asymptotically \cite
{wilkins,quinn}, but the $\ln T$ correction is almost indistinguishable from
$T^2$ at low $T$.

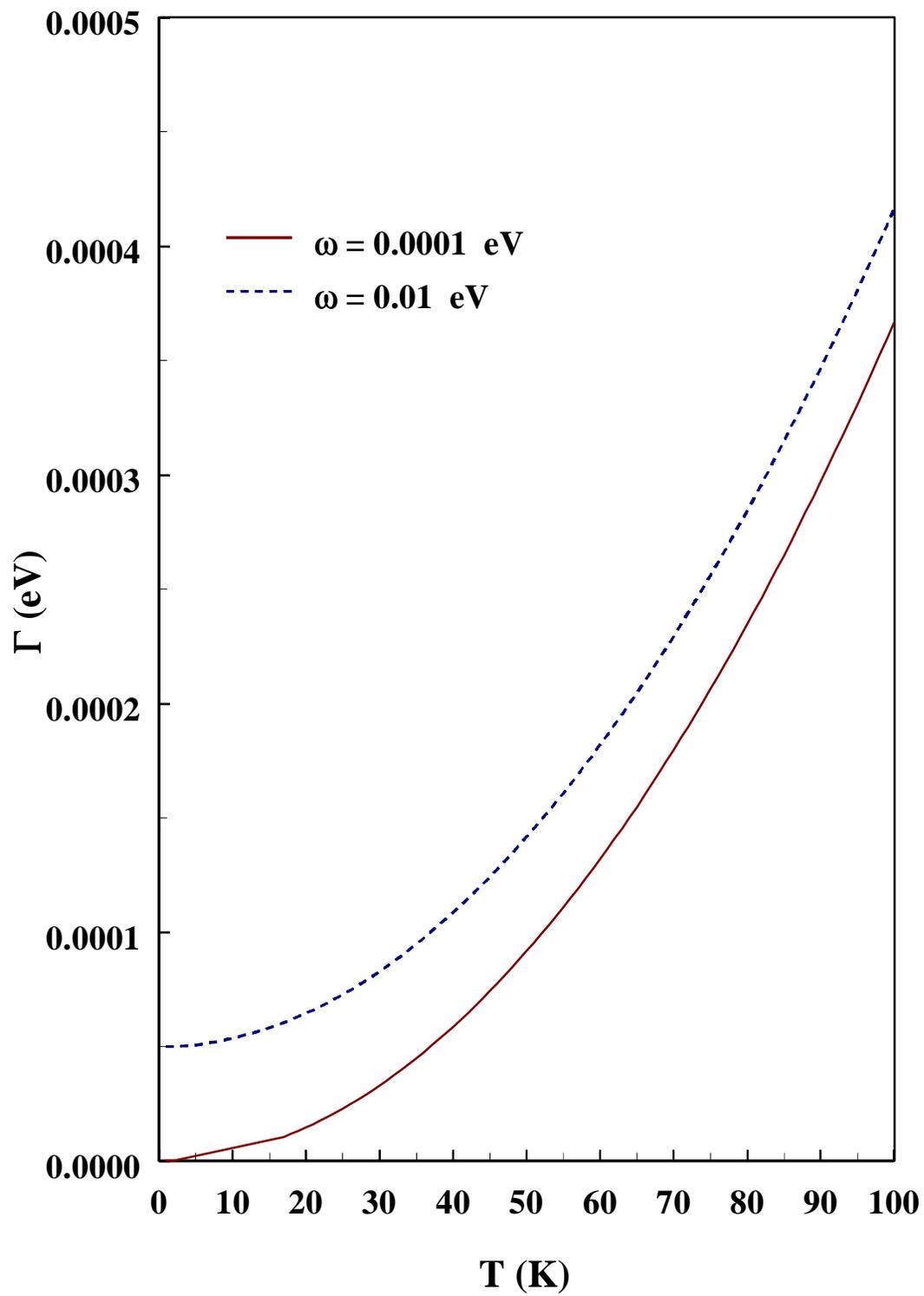
\begin{figure}
\caption{The normal state quasiparticle damping $\Gamma(\omega)$ ({\it i.e.}
the imaginary part of the self energy) for a two dimensional Fermi liquid is
plotted as a function of temperature at two different frequencies, $%
\omega=0.01$ eV and $\omega=10^{-4}$ eV. The standard values of the coupling
$UN(0)=1$ and bandwidth $W=1$ eV produce the small magnitude of $\Gamma$.}
\label{normg}
\end{figure}

The quadratic temperature variation of our computed FL damping at two
different frequencies is shown in Figure \ref{normg} for a coupling $UN(0)=1$
and a bandwidth $W=1$ eV. These parameters are used throughout this work. As
expected, the value of the FL damping seen in Figure \ref{normg} is quite
small. For comparison, the experiments that we examine suggest much larger
estimates of the damping which may perhaps indicate anisotropic Fermi
surface contributions, RPA enhancement of the spin susceptibility, or higher
order scattering. Also, the Hubbard model restricts the Coulomb interaction
to a point in real space and thus samples only the zero angular momentum
scattering channel.

\subsection{Superconducting State}

In the absence of the self energy corrections from particle collisions, the
BCS susceptibility in the superconducting state is modified by the presence
of coherence factors due to an energy gap, and thus becomes
\begin{equation}
\chi \left( {\bf q},\omega \right) =\sum_{{\bf k}}\left[ \text{A}_{{\bf k,q}%
}^{+}\frac{f\left( E_{{\bf k+q}}\right) -f\left( E_{{\bf k}}\right) }{\omega
-E_{{\bf k+q}}+E_{{\bf k}}+i\delta }+\frac 12\text{A}_{{\bf k,q}}^{-}\frac{%
1-f\left( E_{{\bf k+q}}\right) -f\left( E_{{\bf k}}\right) }{\omega +E_{{\bf %
k+q}}+E_{{\bf k}}+i\delta }+\frac 12\text{A}_{{\bf k,q}}^{-}\frac{f\left( E_{%
{\bf k+q}}\right) +f\left( E_{{\bf k}}\right) -1}{\omega -E_{{\bf k+q}}-E_{%
{\bf k}}+i\delta }\right]  \label{schi}
\end{equation}
where the coherence factors are
\begin{equation}
\text{A}_{{\bf k,q}}^{\pm }=\frac 12\left[ 1\pm \frac{\epsilon _{{\bf k}%
}\epsilon _{{\bf k+q}}+\Delta _{{\bf k}}\Delta _{{\bf k+q}}}{E_{{\bf k}}E_{%
{\bf k+q}}}\right]
\end{equation}
and $\Delta _{{\bf k}}$ is the superconducting energy gap, $\epsilon _{{\bf k%
}}$ the energy dispersion in the normal state, and $E_{{\bf k}}=\sqrt{%
\epsilon _{{\bf k}}^2-\Delta _{{\bf k}}^2}$ the quasiparticle energy in the
superconducting state. All the calculations described in this article are
carried out using a superconducting gap of s-wave symmetry. The temperature
dependence of the gap is taken to have the standard form that fits the
solution of the BCS weak coupling gap equation,
\begin{equation}
\Delta (T)=\Delta _0\tanh \left( {1.76\sqrt{{\frac{T_c}T}-1}}\right)
\end{equation}
where $\Delta _0=1.76k_BT_c$. In the limit of negligible damping, $\delta
\rightarrow 0^{+}$, the above susceptibility (Eq. \ref{schi}) gives the
standard sharp Hebel--Slichter peak in the NMR response.

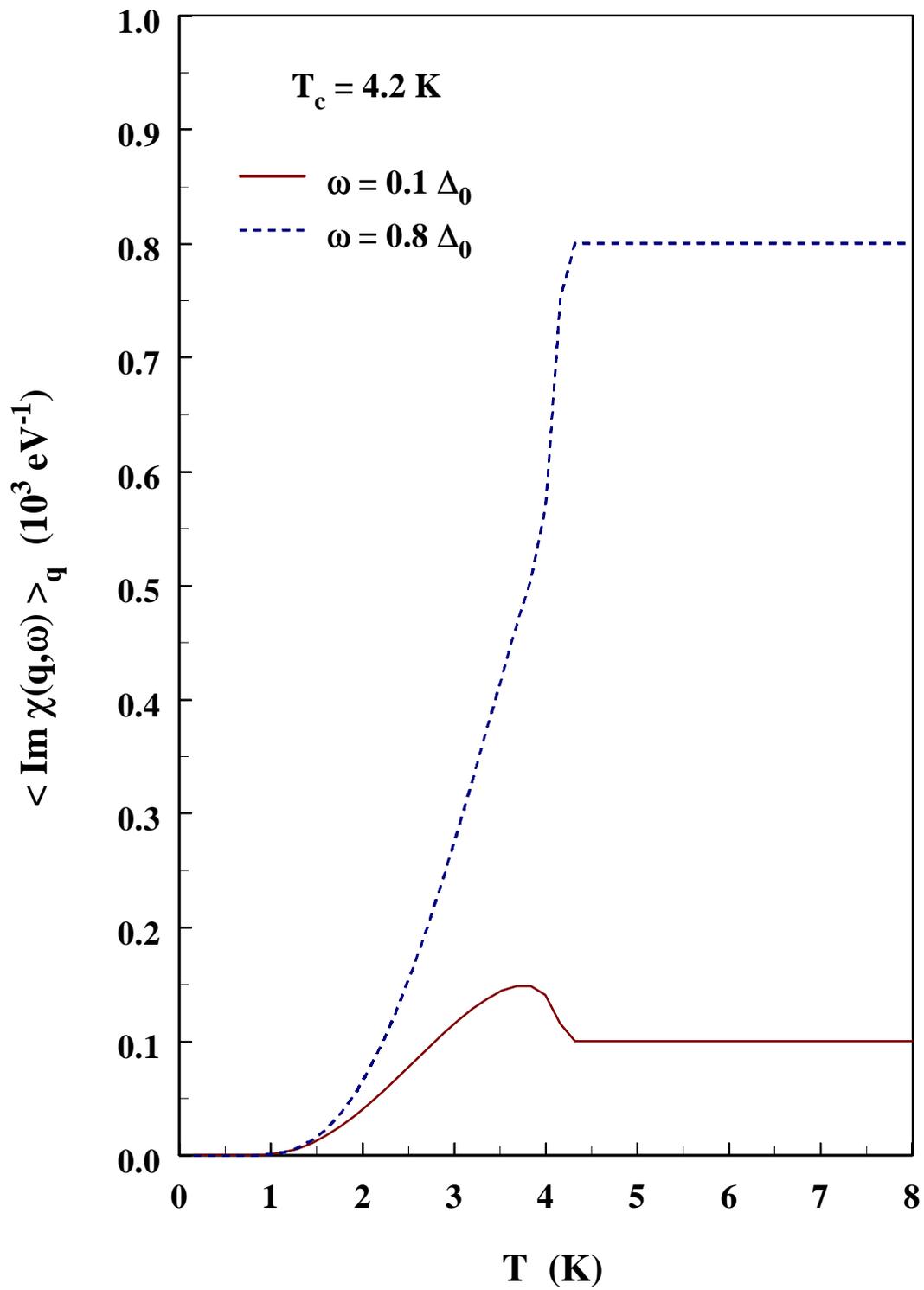
\begin{figure}
\caption{The momentum averaged BCS susceptibility in the superconducting
state is plotted as a function of temperature at two frequencies, $%
\omega=0.1\Delta _0$ and $\omega =0.8\Delta _0$, for a transition
temperature $T_c=4.2$ K. Note that the overall decrease in $\langle
\chi^{\prime \prime}\rangle _q$ at very low $T$ remains a feature at all
frequencies, while the coherence peak is only visible at lower frequencies.}
\label{chi}
\end{figure}

The quasiparticle damping due to spin fluctuations in the superconducting
state is \cite{scdamp}
\begin{eqnarray}
\Gamma \left( {\bf p},\omega \right) &=&\frac{U^2}{\left[ 1-f\left( \omega
\right) \right] }\int \frac{d^2p^{\prime }}{\left( 2\pi \right) ^2}%
\mbox{\LARGE \{ }\int_0^{\omega -\Delta }d\Omega \chi ^{\prime \prime
}\left( {\bf p-p}^{\prime },\Omega \right) \delta \left( \omega -\Omega -E_{%
{\bf p}^{\prime }}\right) \left[ 1+\frac{\Delta ^2}{\omega \left( \omega
-\Omega \right) }\right] \left[ n\left( \Omega \right) +1\right] \left[
1-f\left( \omega -\Omega \right) \right]  \nonumber \\
&&\ \ \ \text{ }+\text{ }\int_{\omega +\Delta }^\infty d\Omega \chi ^{\prime
\prime }\left( {\bf p-p}^{\prime },\Omega \right) \delta \left( \Omega
-\omega -E_{{\bf p}^{\prime }}\right) \left[ 1-\frac{\Delta ^2}{\omega
\left( \Omega -\omega \right) }\right] \left[ n\left( \Omega \right)
+1\right] f\left( \Omega -\omega \right)  \nonumber \\
&&\ \ \ \text{ }+\text{ }\int_0^\infty d\Omega \chi ^{\prime \prime }\left(
{\bf p-p}^{\prime },\Omega \right) \delta \left( \Omega +\omega -E_{{\bf p}%
^{\prime }}\right) \left[ 1+\frac{\Delta ^2}{\omega \left( \Omega +\omega
\right) }\right] n\left( \Omega \right) \left[ 1-f\left( \Omega +\omega
\right) \right] \mbox{\LARGE \} }  \label{gammasc}
\end{eqnarray}
where $n(\omega )$ and $f(\omega )$ are the Bose and Fermi functions
respectively. To simplify the multiple integrations needed for the damping,
we note that the momentum variation of the susceptibility arising from a
free electron dispersion is relatively smooth, in contrast to tight binding
models considered, for example, by Quinlan {\it et al.}\cite{quinlan}. Hence
we approximate the damping calculation by taking a momentum average of the
susceptibility (for $\omega >0$)
\begin{eqnarray}
\left\langle \chi ^{\prime \prime }\left( {\bf k},\omega \right)
\right\rangle _q &=&\frac{\pi N(0)}W\mbox{\LARGE [ }2\int_\Delta ^\infty dE%
\frac{E(\omega +E)+\Delta ^2}{\sqrt{E^2-\Delta ^2}\sqrt{\left( \omega
+E\right) ^2-\Delta ^2}}\left[ f(E)-f(\omega +E)\right]  \nonumber \\
&&\ \ \text{ }+\text{ }\int_\Delta ^{\omega -\Delta }dE\frac{E(\omega
-E)-\Delta ^2}{\sqrt{E^2-\Delta ^2}\sqrt{\left( \omega -E\right) ^2-\Delta ^2%
}}\left[ 1-f(E)-f(\omega -E)\right] \mbox{\LARGE ] }  \label{chiav}
\end{eqnarray}
The resulting calculated susceptibility average is shown in Figure \ref{chi}%
, where the peak just below $T_c$ at low frequencies (solid line) is a
consequence of the coherence factors. At higher frequencies, as seen in the
case of $\omega =0.8\Delta _0$ (dashed line), the susceptibility drops off
rapidly as $T$ decreases though there is no coherence peak below $T_c$. This
characteristic drop of the susceptibility below $T_{c\text{ }}$is the key to
determining the temperature variation of the quasiparticle damping in the
superconducting state. We have used here a superconducting transition
temperature $T_c=4.2$ K, which corresponds to the organic compound (MDT-TTF)$%
_2$AuI$_2$.

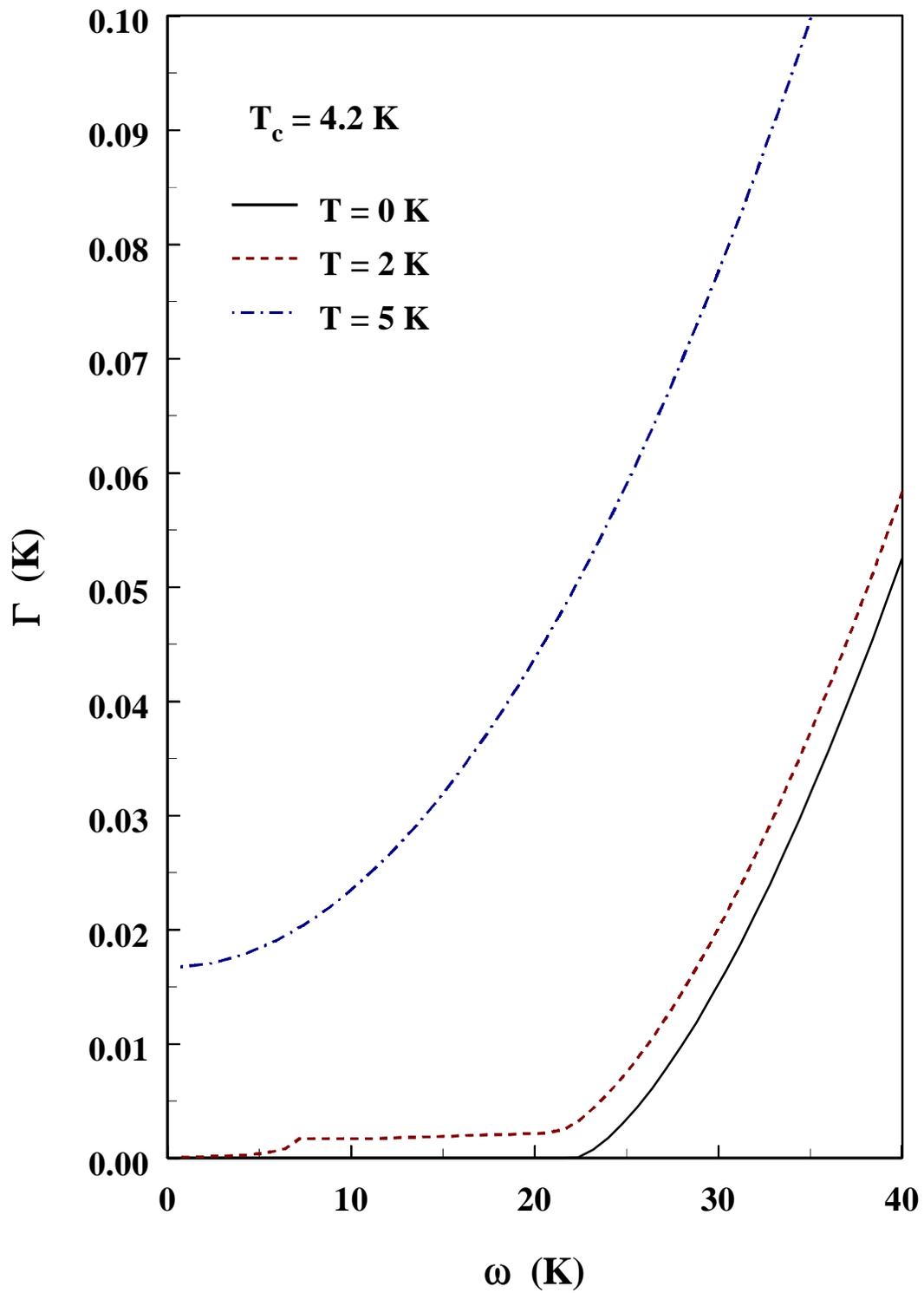
\begin{figure}
\caption{ The calculated quasiparticle damping in the superconducting state
is plotted as a function of frequency at three different temperatures. We
have used here the parameters $UN(0)=1$, $W=1$ eV and $T_c=4.2$ K. Note that
at $T=0$ the damping vanishes below $\omega =3\Delta _0$. At higher
temperatures, however, there is a finite response below $3\Delta (T)$ with
an unusual frequency variation. The damping is quadratic in frequency for $%
\omega>3\Delta (T)$. }
\label{gamw}
\end{figure}

We next compute the quasiparticle damping in Eq. (\ref{gammasc}) as a
function of frequency and temperature for a superconductor with $T_c=4.2$ K.
The frequency dependence of $\Gamma $ is shown in Figure \ref{gamw} at three
different temperatures, and it reveals the dramatic drop in $\Gamma $ at low
$\omega $ caused by the isotropic energy gap. Note that at zero temperature
there is no structure in $\Gamma $ below $3\Delta _0$. At higher
temperatures, $\Gamma $ displays an unusual frequency variation below $%
3\Delta (T),$ and displays a roughly quadratic increase at higher
frequencies.

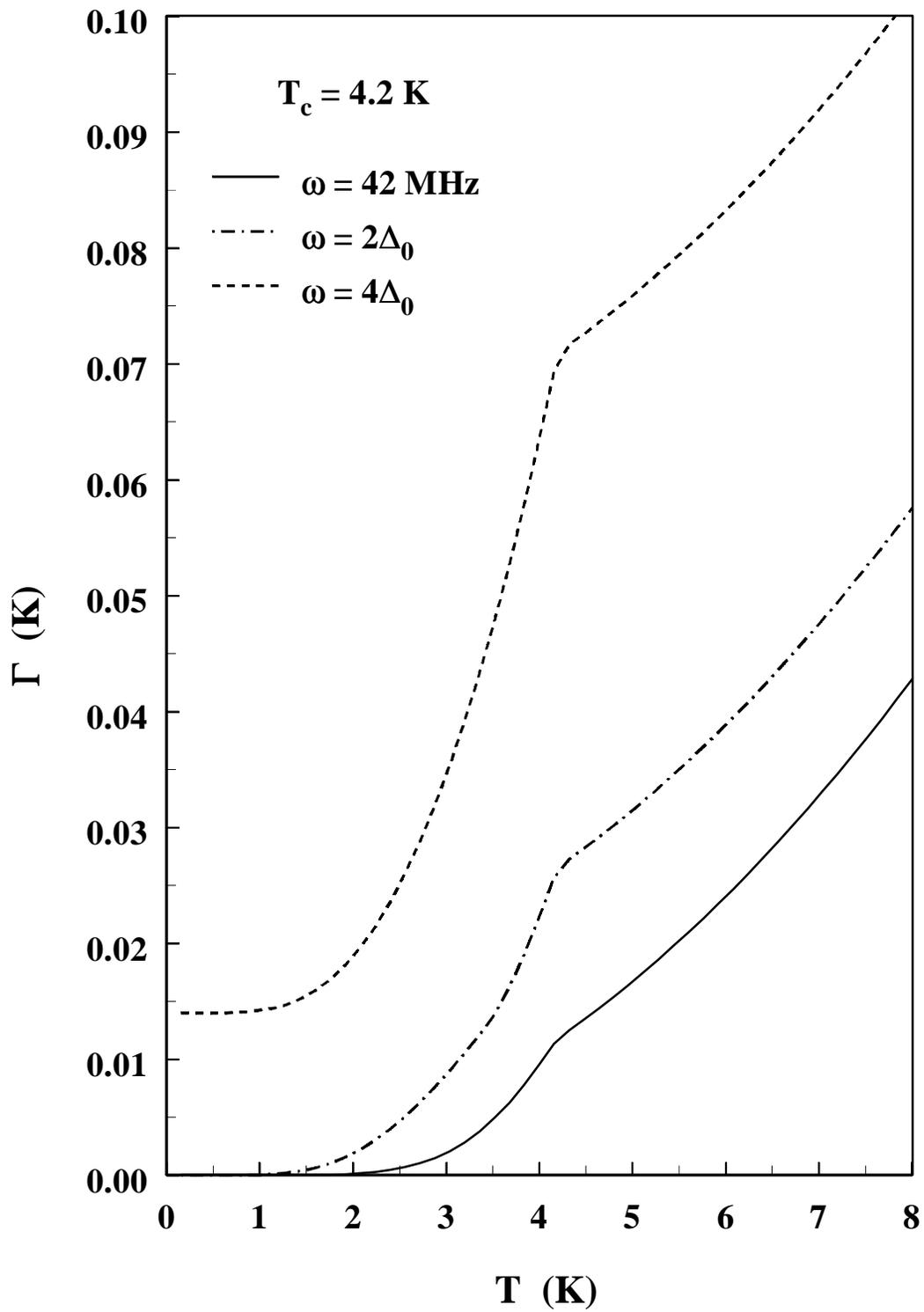
\begin{figure}
\caption{ The calculated quasiparticle damping in the superconducting state
is plotted as a function of temperature at three different frequencies. We
use the transition temperature $T_c=4.2$ K, and $UN(0)=1$, $W=1$ eV. In the
normal state, the damping is quadratic in temperature. As T decreases below
the superconducting transition, the damping drops rapidly to zero at low
frequencies ({\it i.e} $\omega\leq 3\Delta_0$) and saturates at a value that
depends on the frequency at higher $\omega $.}
\label{gamt}
\end{figure}

The temperature dependence of the computed $\Gamma $ is shown in Figure \ref
{gamt} at three different frequencies. For frequencies below $\omega
=3\Delta _0$, the damping drops to zero at low $T$, but reduces to a finite
value even in the zero temperature limit when the frequency exceeds thrice
the energy gap.

For the calculation of the NMR relaxation rate described in the next
section, we reduce the computational complexity by developing a model fit to
the numerically calculated frequency and temperature dependence of the Fermi
Liquid $\Gamma $ at a given value of $T_c$. Above $T_c$, the damping is
presumed quadratic in frequency and temperature.

\section{Nuclear Relaxation Rate}

The standard expression for the NMR relaxation in normal metals is
\begin{equation}
{\frac 1{T_1T}}=|A|^2{\frac 1{\omega _0}}\sum_q\chi ^{\prime \prime
}(q,\omega _0)  \label{t1n}
\end{equation}
where $|A|$ denotes the hyperfine coupling and $\omega _0\sim 10$ MHz is the
typical radio frequency of the measurement. For a Fermi Liquid, the momentum
average of the spin susceptibility in Eq. (\ref{chifl}) is linear in
frequency, and thus produces the familiar Korringa relation for the nuclear
relaxation that is of the form
\begin{equation}
{\frac 1{T_1T}}\sim |A|^2N^2(0)={\rm constant}.
\end{equation}

If the self energy is only weakly momentum dependent, the general form of
the susceptibility in Eq. (\ref{t1n}) can be averaged over $k$ and $q$
independently, and the Korringa behavior persists with very little
contribution from the damping in the normal state. However, an isotropic
energy gap produces a divergent density of states in the superconducting
state which is naturally quite sensitive to the form of the damping. Since
the above analysis demonstrates that the structure of the damping is
particularly important at frequencies comparable to the energy gap, these
features are important for the NMR spectra in the superconducting regime.

\subsection{Self Energy Corrections}

The NMR relaxation rate is computed using the form\cite{schrieffer},
\begin{equation}
\frac 1{T_1}=2\left| A\right| ^2N^2(0)\int_\Delta ^\infty \frac{EdE}{\sqrt{%
E^2-\Delta ^2}}\int_\Delta ^\infty \frac{E^{\prime }dE^{\prime }}{\sqrt{%
E^{\prime 2}-\Delta ^2}}\left( 1+\frac{\Delta ^2}{EE^{\prime }}\right)
f(E)\left[ 1-f(E^{\prime })\right] \frac \Gamma {(\omega +E-E^{\prime
})^2+\Gamma ^2}  \label{rate}
\end{equation}
where $N(0)$ is the normal state density of states at the Fermi energy, and
the spin hyperfine coupling $\left| A\right| $ is taken to be constant,
thereby neglecting crystalline anisotropy. The delta function in the
original expression for $\chi ^{\prime \prime }$ in Eq. (\ref{t1n}) is
replaced by a Lorentzian of width $\Gamma $, the quasiparticle damping
computed in Eq. (\ref{gammasc}). Thus the conventional BCS result is
recovered in the limit $\Gamma \rightarrow 0$. The self energy corrections
included in Eq. (\ref{rate}) depend on both temperature and energy, and
enter as $\Gamma =\frac 12\left[ \Gamma (E,T)+\Gamma (E^{\prime },T)\right]$%
. In all the calculations discussed below, the damping is defined as
\begin{equation}
\Gamma =\alpha \Gamma _{\text{FL}}
\end{equation}
where $\alpha $ is varied to show the consequences of enhancements beyond
the Born approximation used here with one standard set of parameter values $%
W=1$ eV and $UN(0)=1$. A remarkable feature of the organic compound
resistivities shown in Figure \ref{rho} is that their high values would
indicate very short mean free paths if a standard transport model is
applied. However, the temperature variation of the resistance is compatible
with Fermi liquid behavior, so that the enhancement remains anomalous in
these cases even though the likely physical origin is electron-electron
scattering. Anisotropic Fermi surface effects or higher order scattering
such as the RPA corrections may be likely suspects for this mystery.

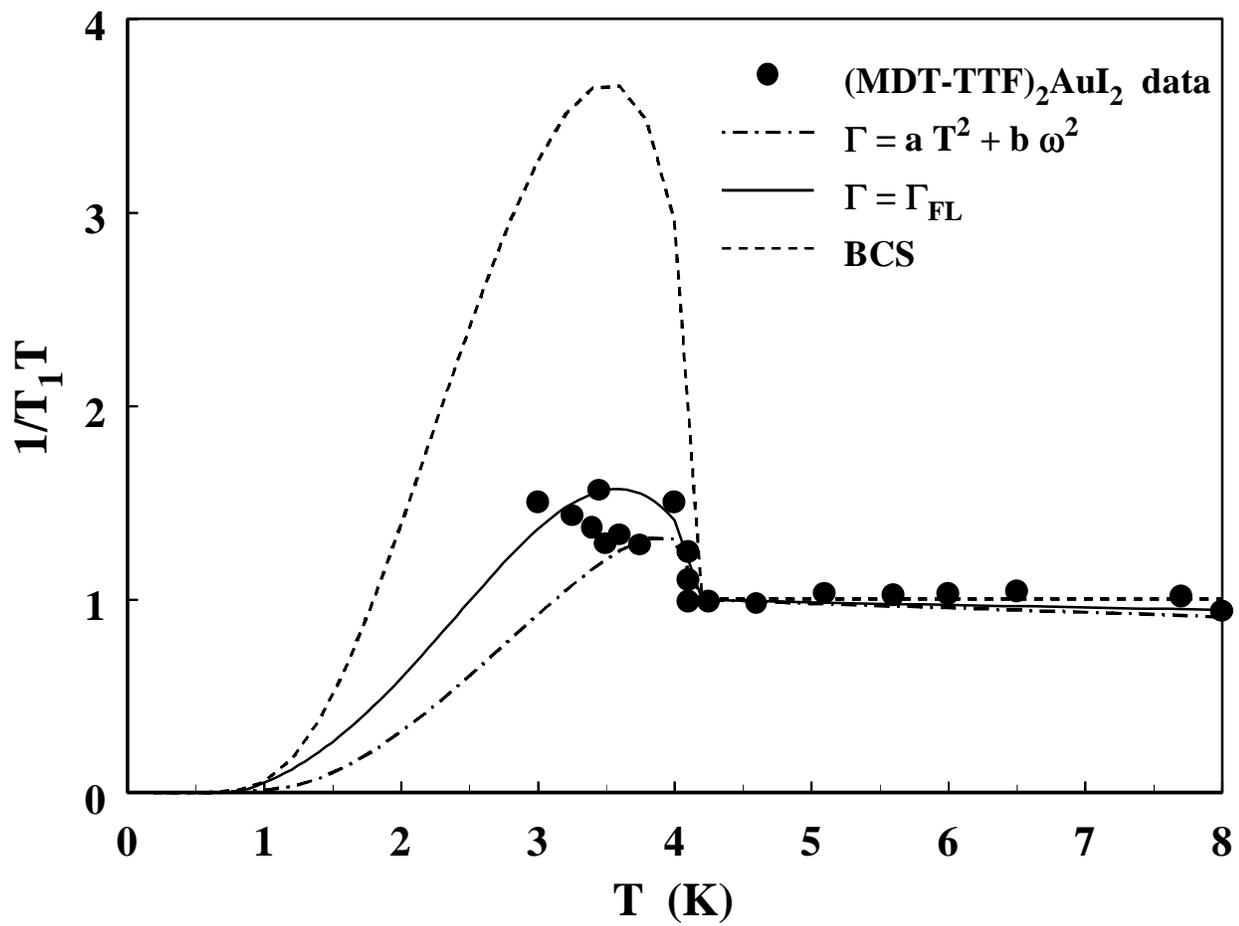
\begin{figure}
\caption{ The NMR relaxation rate $1/T_1T$ data of Takahashi {\it et al.} on
the organic superconductor (MDT-TTF)$_2$AuI$_2$ are plotted as a function of
temperature at a frequency $\omega=42$ MHz. The data show a weak
Hebel-Slichter peak below $T_c=4.2 K$. Also shown are the theoretical curves
for $1/T_1T$ using the calculated Fermi liquid quasiparticle damping $%
\Gamma_{\text{FL}}$ that includes the energy gap (solid line), and a normal
state damping that is quadratic in frequency and temperature at all $T$ and $%
\omega $ (dot-dashed line). The dashed line represents the conventional BCS
result in the limit $\Gamma\rightarrow 0$. In order to fit the (MDT-TTF)$_2$%
AuI$_2$ data, the amplitude of the Fermi Liquid damping $\Gamma_{\text{FL}}$
is increased by a factor of $\alpha=40$. The calculated curves as well as
the data have been normalized to the value of $1/T_{1}T$ at $T_c$.}
\label{mdtttf}
\end{figure}

The relaxation rate data of Takahashi {\it et al.}\cite{takahashi} on the
organic superconductor (MDT-TTF)$_2$AuI$_2$ is shown in Figure \ref{mdtttf}
along with the theoretical curves calculated using Eq. (\ref{rate}). The
solid curve is calculated using the superconducting state quasiparticle
damping computed above (Eq. \ref{gammasc}) for the Fermi liquid. To examine
the influence of the energy gap on the relaxation rate, we compare this to
the dot-dashed curve calculated using the normal-state damping that is
quadratic in frequency and temperature at all $T$ and $\omega $. Note that
in both these cases the Hebel Slichter peak is strongly reduced over the BCS
result (dashed line) obtained in the absence of a broadening $\Gamma $.
However, in order to fit the (MDT-TTF)$_2$AuI$_2$ data, the amplitude
enhancement of the calculated Fermi Liquid damping needs to be $\alpha =40$.
The standard case with $\alpha =1$ resembles the dashed BCS curve because
the damping is very small.

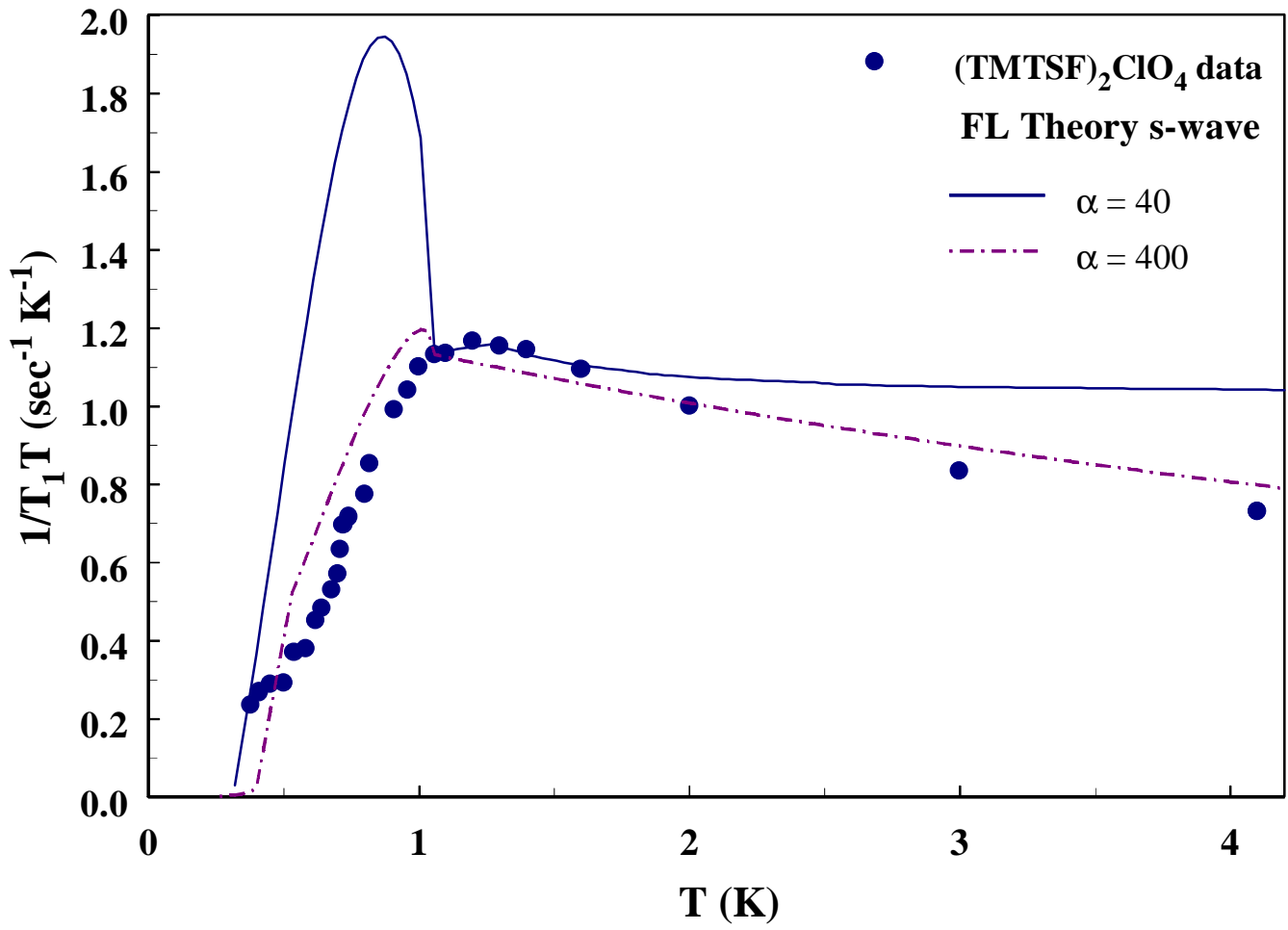
\begin{figure}
\caption{The NMR relaxation rate $1/T_1T$ data of Takigawa {\it et al.} on
the organic superconductor (TMTSF)$_2$ClO$_4$ is shown (dots) as a function
of temperature at a frequency $\omega=42$ MHz. This material has a
transition temperature $T_c=1.06$K and does not display a Hebel-Slichter
peak below $T_c $. Also shown are the theoretical curves for $1/T_1T$ using
the calculated Fermi liquid quasiparticle damping $\Gamma_{\text{FL}}$ where
the magnitude of the damping is increased over the calculated value by a
factor $\alpha$. The curves are shown for two different values of $\alpha$.
The theoretical curves have been normalized to the value of the $1/T_1T$
data at $T=T_c$. Note that even when $\alpha$ is made unrealistically large,
the coherence peak is present in the calculations.}
\label{tmtsf}
\end{figure}

The case of (TMTSF)$_2$ClO$_4$ presents an interesting challenge since the
NMR data of Takigawa {\it et al.} \cite{takigawa} clearly show the absence
of a Hebel--Slichter peak as seen in Figure \ref{tmtsf}. The resistivity of
this organic superconductor has a temperature dependence that is close to $%
T^2$ in some samples \cite{bechgaard}, while carefully quenched samples
exhibit a linear $T$ resistivity\cite{schwenk}. This material is notable for
nesting of the Fermi surface which gives rise to spin density wave (SDW)
phase transitions in a magnetic field as discussed in terms of the nested
orbit quantization by Gorkov and Lebed\cite{gorkov}. Our goal is to fit the
NMR data using the Fermi liquid damping to see whether the damping alone can
account for the suppression of the Hebel--Slichter peak. The results shown
in Figure \ref{tmtsf} reveal that the Fermi liquid damping actually gives a
weak, but nevertheless significant peak in the NMR relaxation just below $%
T_c $ even for very large damping cross sections. The two curves in Fig. \ref
{tmtsf} are calculated using an increase in amplitude of the damping by
factors of $\alpha =$ $40$ and 400 respectively, and both curves are
normalized to the value of $1/T_1T$ at $T_c=1.06$ K. The $\alpha =$ $400$
curve resembles the data, indicating that an anomalously large damping may
indeed be the cause of the suppression of the HS peak.

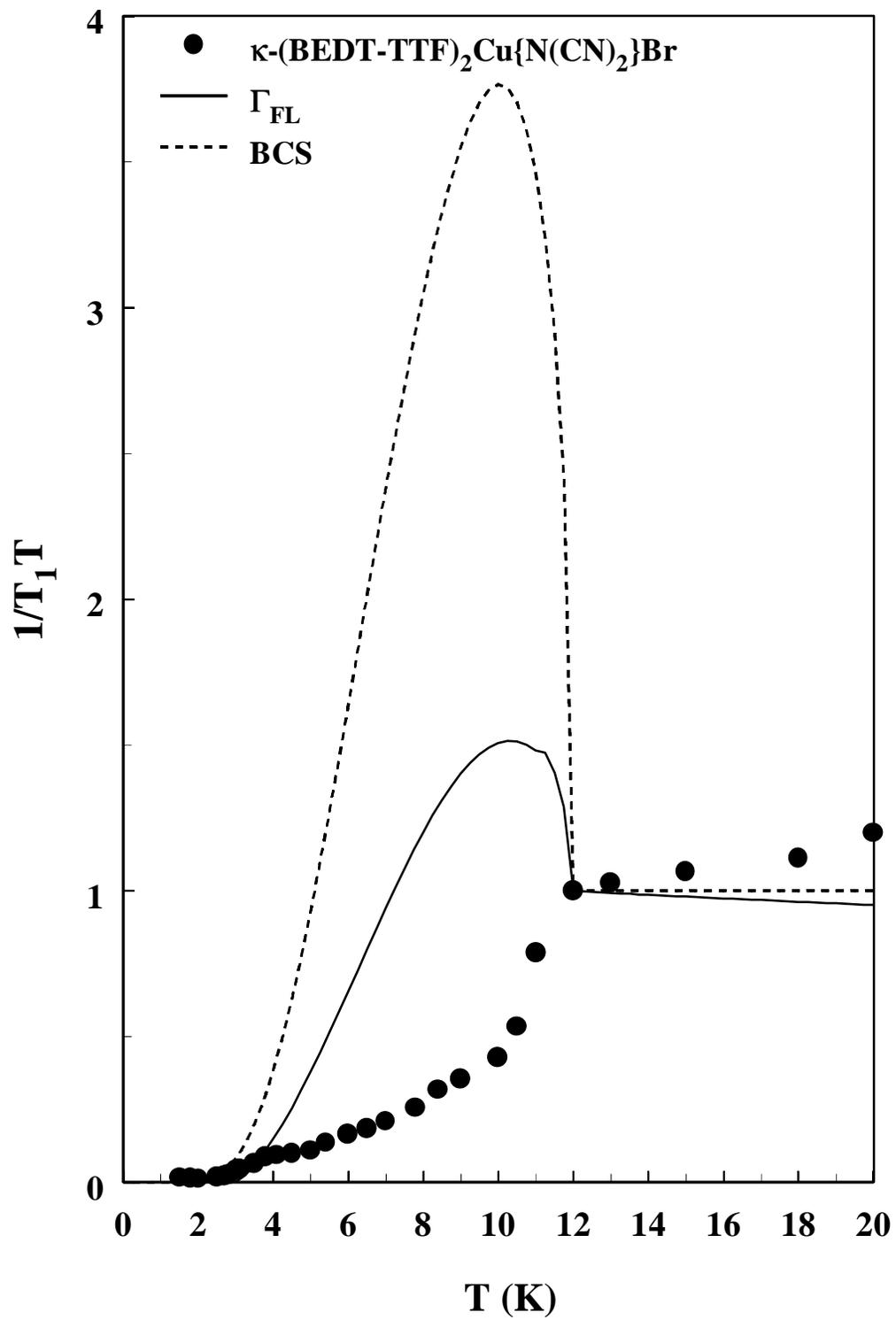
\begin{figure}
\caption{The NMR relaxation rate $1/T_1T$ data of Kanoda {\it et al.} on the
organic superconductor $\kappa -$(BEDT-TTF)$_2$CuN(CN)$_2$Br is plotted as a
function of temperature at a frequency $\omega=25$ MHz. The solid line
represents the theoretical curve for $1/T_1T$ using the calculated Fermi
liquid quasiparticle damping $\alpha\Gamma_{\text{FL}}$ where the magnitude
of the damping is increased over the bare value by a factor $\alpha=40$.
Also shown is the BCS result (dashed line) for which the coherence peak is
more pronounced. The theoretical curves as well as the data have been
normalized to the value of $1/T_1T$ at $T_c=12$ K. The low temperature $T^3$
variation of the data resembles d-wave pairing behavior, and the pressure
variation of the normal state $T_1T$ may indicate spin fluctuation
contributions which are enhanced by Fermi surface nesting, that is beyond
the scope of the present analysis.}
\label{bedt}
\end{figure}

The data shown in Figure \ref{bedt} provide another illustration of
non-Fermi Liquid behavior in the organic compound $\kappa -$(BEDT-TTF)$_2$%
CuN(CN)$_2$Br. The NMR relaxation rate data of Kanoda {\it et al.}\cite
{kanoda} on this organic compound are shown (dots) along with the
theoretical curve (solid line) calculated using the Fermi Liquid damping and
enhancement $\alpha =40$. Unlike the calculated curve, the data show no
evidence of a Hebel-Slichter peak, and also drop much more sharply at low
temperatures than the calculated curves. The conventional BCS result with a
more pronounced coherence peak is also shown (dashed line). The data for
this organic compound exhibit a $T^3$ temperature dependence at low $T$,
which is suggestive of an unconventional pairing state with line nodes on
the Fermi surface, {\it e.g.} d--wave pairing. In the normal state the data
are also anomalous in that they do not follow a Korringa law\cite{bedtnor}.
A peak in $1/T_1$ is observed at 50 K which tends to vanish under pressure%
\cite{mayaffre}. This pressure variation of $T_1T$ in the normal state may
arise from spin fluctuations whose contributions increase with the degree of
nesting of the Fermi surface.

\subsection{Vertex Corrections}

The NMR relaxation rate discussed above has been calculated in the presence
of self energy corrections to the spin susceptibility. Schrieffer \cite{ward}
has raised the issue of the importance of vertex corrections for pairing
schemes that rely on the exchange of spin fluctuations, suggesting that the
pairing coupling may be strongly suppressed. We have addressed this question
in the case of the lowest order bubble susceptibility $\chi (q,\omega )$. We
are interested in vertex corrections to the NMR relaxation rate consistent
with the Ward identity. Since $1/T_1T=\lim_{\omega \rightarrow 0}\sum_q\chi
^{\prime \prime }(q,\omega )/\omega $, it suffices to calculate the
correction to the momentum averaged imaginary part of the spin
susceptibility. Wermbter and Tewordt \cite{vertex} have used strong coupling
Eliashberg theory to compute the vertex corrections within a 2D Hubbard
model for an interaction arising from the exchange of spin and charge
fluctuations as well as phonons. In their calculations, they approximate the
bubble susceptibility $\chi (q,\omega )$ by its average over momentum. Their
approach is thus a good starting point for our calculation which is carried
out in the weak coupling limit for the lowest order electron-hole bubble.

In the absence of self-energy and vertex corrections, the momentum average
of the imaginary part of the susceptibility is given by the expression for $%
1/T_1T$ in Eq. (\ref{rate}). Including vertex corrections to order $U^2$,
the one-loop contribution relevant to the spin fluctuations becomes\cite
{vertex}
\begin{eqnarray}
\bar \chi ^{\prime \prime }(\omega ) &=&\pi \int_{-\infty }^\infty dE\frac{%
E(E+\omega )+\Delta ^2}{\sqrt{E^2-\Delta ^2}\sqrt{(E+\omega )^2-\Delta ^2}}
\nonumber \\
&&\ \ \ \ \ \ \left[ f(E)[1-J_1(E,\omega )+\pi ^2J_2(E,\omega )]-f(E+\omega
)[1-J_1(E,\omega )+\pi ^2J_2^{\prime }(E,\omega )]\right]   \label{vert}
\end{eqnarray}
where $\bar \chi ^{\prime \prime }=\langle \chi ^{\prime \prime }({\bf q}%
,\omega )\rangle _q$ is the momentum-averaged susceptibility. The correction
terms $J_1$, $J_2$ are given by
\begin{eqnarray}
J_1(E,\omega ) &=&-\frac{U^2N^2(0)}{2\pi }\int_{-\infty }^\infty d\nu \bar
\chi ^{\prime \prime }(\nu )\int_{-\infty }^\infty d\mu _1\frac{f(-\mu
_1)+n(\nu )}{\sqrt{\mu _1^2-\Delta ^2}}\int_{-\infty }^\infty \frac{d\mu _2}{%
\omega -\mu _2}  \nonumber \\
&&\ \ \ \ \left\{ \frac{|\mu _1||\mu _1-\mu _2|+\Delta ^2}{(E+\omega -\mu
_1-\nu )\sqrt{(\mu _1-\mu _2)^2-\Delta ^2}}-\frac{|\mu _1||\mu _1+\mu
_2|+\Delta ^2}{(E-\mu _1-\nu )\sqrt{(\mu _1+\mu _2)^2-\Delta ^2}}\right\}
\label{j1}
\end{eqnarray}
where $f(x)$, $n(x)$ are the Fermi and Bose functions respectively, and
\begin{equation}
J_2(E,\omega )=-\frac{U^2N^2(0)}{2\pi }\int_{-\infty }^\infty d\nu \bar \chi
^{\prime \prime }(\nu )[f(\nu -E-\omega )+n(\nu )]\ \ \frac{|E+\omega -\nu
||E-\nu |+\Delta ^2}{\sqrt{(E+\omega -\nu )^2-\Delta ^2}\sqrt{(E-\nu
)^2-\Delta ^2}}.  \label{j2}
\end{equation}
The term $J_2^{\prime }$ in Eq. (\ref{vert}) is obtained from $J_2$ by
substituting $f(\nu -E)$ for $f(\nu -E-\omega )$.

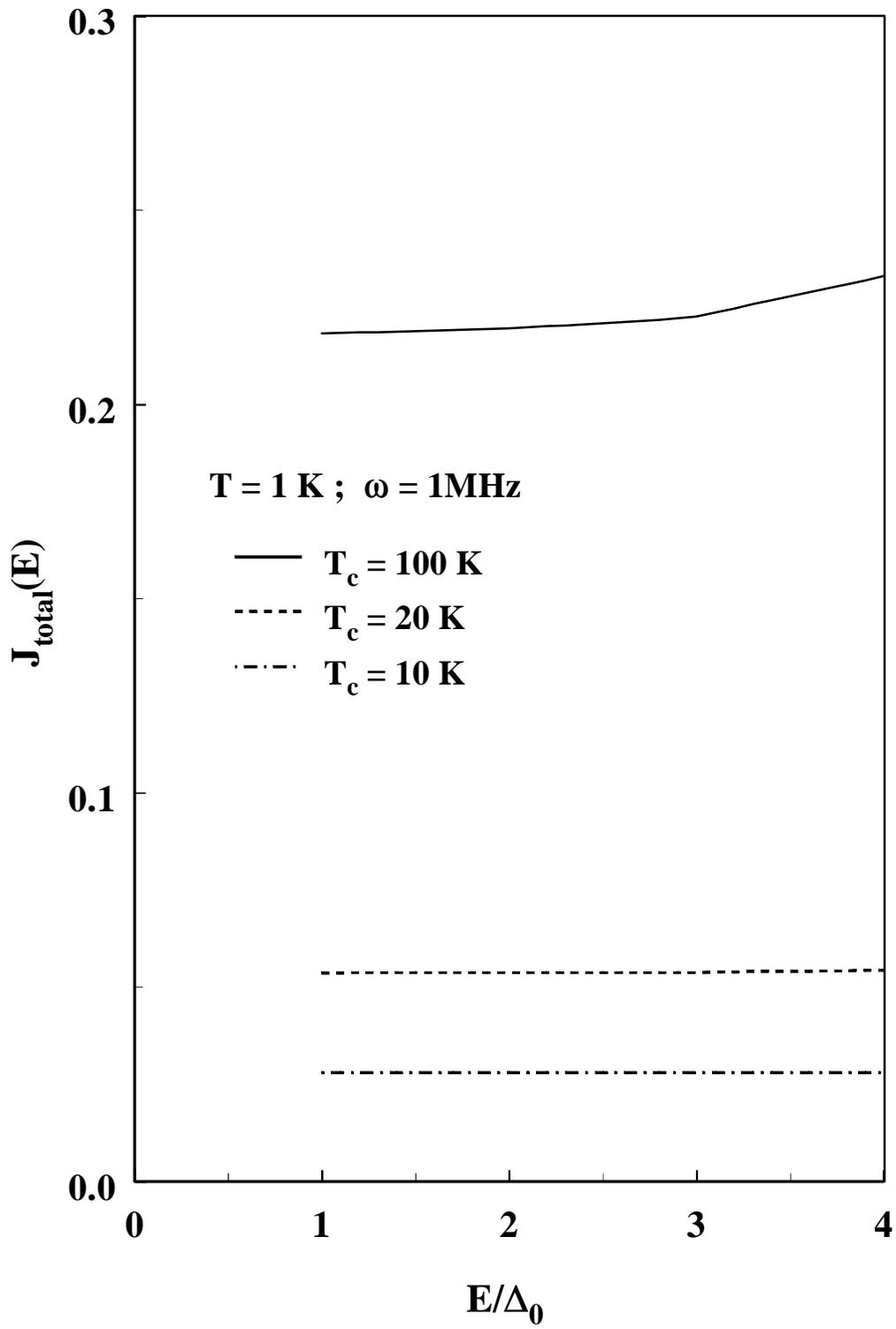
\begin{figure}
\caption{The total contribution $J(E)$ from the calculated vertex
corrections to the NMR relaxation rate is plotted as a function of the
energy $E$ scaled to the energy gap $\Delta$ at a fixed temperature $T=1$K.
The three curves correspond to superconducting transition temperatures of $%
T_c=$ 100, 20 and 10 K. The correction increases with $T_c$ as seen in the
figure. For a fixed $T_c$ the vertex contribution at the $T=1$ K value
chosen here is close to the maximum and falls off as the temperature
approaches $T_c$. For $T_c=10$ K, the magnitude of the correction at $%
E=4\Delta$ is $<3$\%, and thus negligible. We also note that since $J(E)$ is
practically independent of $E$, the effect of this correction is to
renormalize the interaction strength $U$.}
\label{vertex}
\end{figure}

The NMR relaxation rate is obtained from the $\omega \rightarrow 0$ limit of
$\bar \chi ^{\prime \prime }(\omega )$ in Eq. (\ref{vert}). We therefore
calculate $J_1(E,0)$ and $J_2(E,0)$, noting that $J_2^{\prime }(E,0)\equiv
J_2(E,0)$. The $\mu _2$ integration in Eq. (\ref{j1}) can be carried out
analytically, giving
\begin{equation}
J_1(E,0)=\frac{U^2N^2(0)}\pi \int_{-\infty }^\infty d\nu \bar \chi ^{\prime
\prime }(\nu )\int_{-\infty }^\infty d\mu _1[\coth {\frac \nu {2T}}+\tanh {%
\frac{\mu _1}{2T}}]{\frac 1{(E-\mu _1-\nu )}}{\frac{\Delta ^2}{(\mu
_1^2-\Delta ^2)}}\ln \left| \frac{\mu _1+\sqrt{\mu _1^2-\Delta ^2}}{\mu _1-%
\sqrt{\mu _1^2-\Delta ^2}}\right| .  \label{j1p}
\end{equation}
In the low temperature limit, $T\rightarrow 0$, we can obtain an estimate of
the vertex contribution by replacing $\coth {(x/2T)}$, and $\tanh {(x/2T)}$
by sgn$({x})$, and the $\mu _1$ integration in Eq. (\ref{j1p}) can also be
carried out analytically. The remaining $\nu $ integrations in $J_1$ and $%
J_2 $ (Eq. \ref{j2}) are carried out numerically. In the limit $\omega
\rightarrow 0$, Eq. (\ref{vert}) becomes
\begin{equation}
{\frac 1{T_1T}}\propto \lim_{\omega \rightarrow 0}{\frac 1\omega }\bar \chi
^{\prime \prime }(\omega )=\pi \int_{-\infty }^\infty dE\frac{E^2+\Delta ^2}{%
E^2-\Delta ^2}(-{\frac{\partial f}{\partial E}})[1-J_1(E,0)+\pi ^2J_2(E,0)]
\label{vert2}
\end{equation}
The resulting total vertex correction, $-J_1(E,0)+\pi ^2J_2(E,0)$ is shown
in Figure \ref{vertex}. We find that the contribution of $J_1$ is
approximately an order of magnitude larger than $J_2$ for $T<T_c$. Since $%
J_1 $ is proportional to $\Delta ^2$, the correction is larger for a system
with higher $T_c$. This is illustrated in Figure \ref{vertex} for $T_c=$ 10,
20, 100 K. We find that $J_1$ decreases with increasing temperature, and
vanishes as $T\rightarrow T_c$ since $\Delta \rightarrow 0$. In this limit, $%
J_2(E,0)\rightarrow -U^2N^2(0)[\pi ^2T^2+E^2]/4W^2$, which is small for $%
E\ll W$. The temperature dependence of $J_2$ reveals that $%
J_2(T<T_c)<J_2(T_c)$ for all $E$. Thus the total correction term decreases
as $T$ increases. The transition temperatures of the organics are at most of
the order of 10 K, and the total vertex correction for $T_c=10$ K is less
than 3\%. We also find from Figure \ref{vertex} that at a fixed temperature,
regardless of the value of $T_c$, the correction does not depend strongly on
$E$. It can thus be taken out of the integral in Eq. (\ref{vert2}) and the
net result is a renormalized value of the interaction strength $U$.

We basically concur with the conclusions of Wermbter and Tewordt \cite
{vertex} who estimated vertex contributions by another method and found them
to be negligible for their choice of parameters.

\section{Conclusions}

We have calculated the quasiparticle damping $\Gamma $ arising from
electron-electron collisions in a two dimensional Fermi Liquid. We have
presented results for the temperature and frequency variation of $\Gamma $
in the normal and the superconducting state with an isotropic energy gap.
The damping is found to be quadratic in temperature and frequency above the
superconducting transition, consistent with the $T^2$ resistivity of the
organic superconductors and the Nd$_{2-\text{x}}$Ce$_{\text{x}}$CuO$_4$
cuprate.

The computed damping follows an unusual frequency variation in the
superconducting state for frequencies $\omega <3\Delta (T)$. The reduction
in available scattering states due to the opening of an isotropic energy gap
causes the quasiparticle damping to drop at temperatures below the
superconducting transition, as seen in our results. On the other hand an
energy gap of d-wave symmetry has nodes and thus ensures the availability of
scattering states even at $T=0$. This would produce a quasiparticle damping
that drops less rapidly as $T$ decreases below $T_c$.

We have also explored the influence of the calculated FL damping on the NMR
relaxation rate. We find that the damping acts to reduce the magnitude of
the Hebel-Slichter peak below $T_c$, but does not eliminate it. Our results
are compared to experimental data on three organic superconductors. In the
case of (MDT-TTF)$_2$AuI$_2$, our results fit the data and reproduce the
weak HS peak seen in the experiment. In the case of (TMTSF)$_2$ClO$_4$, the
data show no HS peak and deviate from our FL results. We find, however, that
if the damping is made anomalously large, our results resemble the data on
(TMTSF)$_2$ClO$_4$, though a small HS peak is still present in the
calculations. Since the HS peak is sensitive to the BCS coherence factors,
the gap symmetry and the quasiparticle damping, the analogous ``coherence
peak'' in the microwave conductivity should provide a further test of our
analysis. Applications of the FL damping to microwave conductivity and
surface resistance measurements on the NCCO cuprate\cite{anlage} reveal that
these probes are also sensitive to the self--energy structure\cite{shub}.

The NMR data on the BEDT organic superconductors exhibit a very rapid
decrease below $T_c$ that is similar to the behavior seen in the high
temperature cuprate superconductors. These cases appear to require
additional mechanisms to explain the lack of a Hebel-Slichter peak as well
as the curvature of the NMR relaxation below $T_c$. Our microscopic analysis
yields a relaxation curve that remains considerably higher than the BEDT
data in the superconducting state even if the Coulomb coupling is
artificially increased to unrealistic values.

The influence of the symmetry of the gap on the relaxation rate may be seen
from the behavior of the density of states. For an s-wave gap, the
superconducting density of states is zero below $\Delta $ and has a square
root singularity at $\Delta $. The density of states for a d--wave gap is
linear at low energies and has only a logarithmic singularity at $\Delta $.
While the Hebel Slichter peak is greatly reduced as a result of replacing a
square root singularity by a logarithmic one, the relaxation rate in the
d--wave case falls off more slowly than the s--wave case as $T$ decreases
below $T_c$. At low temperatures, the d--wave relaxation rate varies as $T^3$%
, and the BEDT data provide evidence for the latter symmetry.

We note here that in the case of the NFL damping calculated for a nested
Fermi surface by Rieck {\it et al.}\cite{rieck}, the quasiparticle damping
drops to zero at frequencies $\omega <3\Delta $ more rapidly than the FL
damping discussed here. Also the nesting generates an enhanced peak in the
susceptibility at the nesting vector whose suppression by an isotropic
energy gap is more pronounced than the case of the averaged susceptibility
that dominates the Fermi liquid response in the present work.

The extension of the present analysis to incorporate a d-wave energy gap is
in progress. Future studies of a model Fermi surface that includes nesting
features are warranted in view of the remaining anomalies in the NMR spectra
of the BEDT organic compound as well as the cuprate superconductors.

\acknowledgments

It is a pleasure to thank Attila Virosztek for helpful discussions, and
David Djajaputra, Carsten Rieck and Jeff Thoma for their input. We thank K.
Kanoda for sending us their preprint and T. Takahashi for useful exchanges.
We also thank K. Kanoda, T. Takahashi and M. Takigawa for use of their data.
This work was supported by DOE Grant No. DEFG05-84ER45113.

\end{document}